\documentclass[aps,prl,twocolumn,amsmath,amssymb,floatfix,longbibliography,superscriptaddress]{revtex4-2}
\usepackage[T1]{fontenc}
\usepackage{blindtext}
\usepackage{lineno}
\usepackage{graphicx}% Include figure files
\usepackage{dcolumn}% Align table columns on decimal point
\usepackage{bm}% bold math
\usepackage[normalem]{ulem} 
\usepackage{amsmath,bm}
\usepackage{amsfonts}
\usepackage{amssymb,enumerate}
\usepackage{graphicx,hyperref}
\usepackage{amssymb}
\usepackage{epstopdf,booktabs}
\usepackage{empheq}
\usepackage{fancybox}
\usepackage{ulem}
\usepackage{upgreek}
\usepackage{amssymb}
\usepackage[usenames,dvipsnames]{color}
\usepackage[utf8]{inputenc}
\usepackage{natbib}
\usepackage{soul}
\hypersetup{hidelinks,colorlinks=true,citecolor=blue,linkcolor=blue,urlcolor=black}

\newcommand*{\rom}[1]{\expandafter\@slowromancap\romannumeral #1@}

\begin{document}
\title{Intense tunable terahertz radiation from phase-matched difference frequency generation in strongly magnetized plasmas}

\author{Sida Cao}
\email{sidacao@stanford.edu}
\affiliation{Department of Mechanical Engineering, Stanford University, Stanford, California 94305, USA}
% \author{D. Singh}
% \affiliation{Department of Mechanical Engineering, Stanford University, Stanford, California 94305, USA}
% \author{Ke Ou}
% \affiliation{Department of Mechanical Engineering, Stanford University, Stanford, California 94305, USA}
% \author{V. M. Perez-Ramirez}
% \affiliation{Department of Mechanical Engineering, Stanford University, Stanford, California 94305, USA}
% \author{M. M. Wang}
% \affiliation{Department of Electrical and Computer Engineering, Princeton University, Princeton, New Jersey 08540, USA}
% \author{J. M. Mikhailova}
% \affiliation{Department of Mechanical and Aerospace Engineering, Princeton University, Princeton, New Jersey 08540, USA}

% \author{P. Michel}
% \affiliation{Lawrence Livermore National Laboratory, Livermore, California 94551, USA}

\author{Matthew R. Edwards}
\email{mredwards@stanford.edu}
\affiliation{Department of Mechanical Engineering, Stanford University, Stanford, California 94305, USA}

\date{\today}

\begin{abstract}
High-energy terahertz pulses are challenging to produce due to the low conversion efficiency and limited optical damage threshold of nonlinear crystals. 
Here, we describe the high-efficiency generation of terahertz radiation pulses with tunable frequency and field strengths exceeding $500\ \mathrm{GV/m}$ by propagating two-color laser pulses through a strongly magnetized plasma. 
The field strength is substantially enhanced by utilizing two extraordinary-mode branches to minimize the phase mismatch.
We derive the phase-matching conditions and characterize the nonlinear coupling analytically, and validate these predictions with particle-in-cell simulations.
These results establish a new pathway toward next-generation intense terahertz sources with performance well beyond the limits of existing plasma mechanisms and conventional crystal-based approaches.
\end{abstract}

\maketitle
Terahertz (THz) radiation with field strengths exceeding hundreds of $\mathrm{GV/m}$ would open access to relativistic light-matter interactions at long wavelengths~\cite{Mourou2006optics}, a regime previously restricted to high-power near-infrared laser systems at micron and sub-micron wavelengths~\cite{liao2023perspectives}. 
Operating in this relativistic long-wavelength regime is expected to produce interaction dynamics distinct from those driven by shorter-wavelength lasers.
For example, intense mid-infrared to THz fields (with wavelengths from several to tens of microns) can enhance particle acceleration~\cite{palfalvi2014evanescent, sharma2016intense,sharma2018terahertz} and frequency conversion~\cite{hernandez2013zeptosecond,liang2017high,thiele2019electron}, and drive strong nonlinear material responses~\cite{qi2009collective}.
In addition, high-energy mid-infrared to terahertz pulses combined with near-infrared fields can enhance high-order harmonic generation and enable spectral control of attosecond pulses~\cite{serrat2010all,balogh2011single,kovacs2012quasi,rumiantsev2025observation}. 
Despite the exciting physics afforded by high-field terahertz pulses, existing generation techniques---optical rectification and parametric wave mixing in nonlinear crystals and gases~\cite{cook2000intense, bartel2005generation, sell2008phase, kim2008coherent, junginger2010single, vicario2014gv, knorr2017phase, liang2023generation, kim2024ionizing}---are limited to field strengths of at most tens of GV/m, corresponding to a normalized vector potential $a_0=eE/m_e\omega c$ of only around 0.1, more than an order of magnitude below the relativistic regime, where $e$ is the elementary charge, $E$ is the maximum electric field of the pulse, $m_e$ is the electron mass, $\omega$ is the light frequency and $c$ is the speed of light. Further increasing the field strength is fundamentally limited by two factors: the low efficiency of the nonlinear conversion process (typically around $10^{-5}$-$10^{-3}$), and optical damage thresholds of nonlinear crystals, which impose an upper bound on the admissible input intensity.

Plasma-based THz generation overcomes the limitations of conventional nonlinear media by exploiting the orders-of-magnitude higher damage threshold of plasmas~\cite{hamster1993subpicosecond,liao2016demonstration, pak2023multi, simpson2024dephasingless, maity2025enhanced}.
In laser-plasma interactions, terahertz radiation can be produced through several distinct mechanisms: linear mode conversion~\cite{sheng2005emission,liao2015bursts,pukhov2021efficient,wang2024millijoule}, plasma dipole radiation~\cite{kwon2018high,lee2023intense,kumar2025narrowband}, plasma filamentation in air~\cite{xie2006coherent,andreeva2016ultrabroad,koulouklidis2020observation}, photoionization-induced radiation and coherent transition radiation~\cite{leemans2003observation,pearson2011simulation,clerici2013wavelength, miao2016laser,miao2017high,dechard2018terahertz}, and photon deceleration in plasma wakefields~\cite{dechard2019thz}.
Some of these mechanisms have successfully produced millijoule pulses of terahertz radiation~\cite{wang2024millijoule}.
Applying an external magnetic field has further been proposed as a route to enhance the radiated pulse energy by exploiting the stronger coupling between the magnetized plasma wake and the terahertz radiation~\cite{yoshii1997radiation, spence2001simulations, yugami2002experimental}.
Strong magnetic fields additionally enable polarization control and topological manipulation of intense terahertz pulses~\cite{wang2015tunable,bogatskaya2020circularly, kumar2021simulation, tailliez2022terahertz, tailliez2023terahertz, cai2025dynamic, cai2025generation}. 
Although these approaches take advantage of the high damage threshold of plasma, the generated terahertz pulses remain limited to a $a_0\approx 0.1$~\cite{tailliez2022terahertz}, with restricted conversion efficiency and limited control over frequency and pulse duration.

In this Letter, we demonstrate that difference frequency generation in a strongly magnetized plasma can efficiently produce terahertz radiation with tunable frequency (1-100 THz), controllable pulse duration (from single- to multi-cycle), and peak field strengths exceeding 500 GV/m, reaching the relativistic regime with $a_0\approx 7$.
High field strength and conversion efficiency are achieved by using two branches of the extraordinary (X) mode to minimize phase mismatch in the wave-mixing process.
The phase-matched terahertz frequency is determined by the frequency difference and mean frequency of the two pump pulses, as well as the phase-matching conditions. 
We derive analytic phase-matching conditions and a model for the nonlinearity of the conversion process, both of which show excellent agreement with particle-in-cell (PIC) simulations. 
Three-dimensional simulations confirm that multidimensional effects have negligible impact on the output field strength. 

\begin{figure}[t]
    \centering
    \includegraphics[width=1\linewidth]{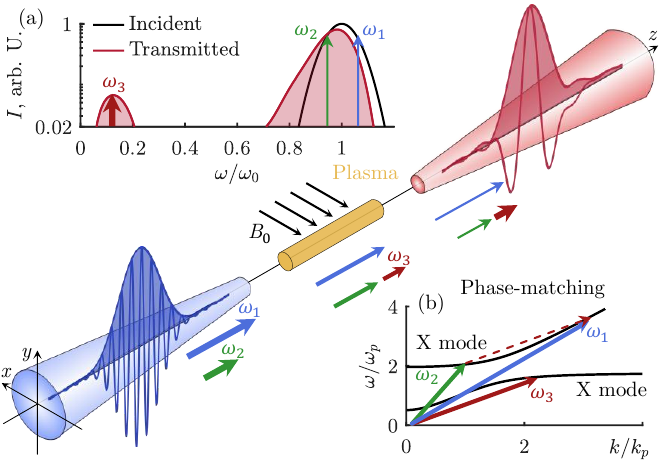}
    \caption{A schematic of THz generation using an ultrashort broadband pump pulse with $\alpha = 0.05$ in a strongly magnetized plasma, with magnetic field $B_c = 0.4$, plasma density $N_0 = 0.01$, and plasma length $L = 100\lambda_0$. The pump is shown in blue, the generated terahertz pulse in red, and the plasma in orange. 
    (a) The spectra of the incident and the transmitted pulse. 
    (b) A dispersion diagram of the phase-matched THz generation process.}
    \label{fig:THz_schematic}
\end{figure}

THz generation via phase-matched difference frequency generation is illustrated in Fig.~\ref{fig:THz_schematic}.
A broadband pump pulse containing frequency components $\omega_1$ and $\omega_2$ propagates through a strongly magnetized plasma, with the external magnetic field oriented orthogonal to both the pump polarization and propagation direction. 
As the pulse propagates, it drives density waves oscillating at $\omega_1$ and $\omega_2$~\cite{Supplemental_Material}, which subsequently interact with the pump to produce a nonlinear current at the difference frequency $\omega_3 = \omega_1 - \omega_2$. 
This current drives difference frequency generation, producing electromagnetic radiation at $\omega_3$, as shown in Fig.~\ref{fig:THz_schematic}. 
For pump pulses in the near-infrared to mid-infrared range, $\omega_3$ falls within the mid-infrared to THz band. 
As an alternative to a single broadband pulse, the two frequency components may belong to two distinct, temporally overlapped pulses.
The spectra of the incident and transmitted pulses are shown in Fig.~\ref{fig:THz_schematic}(a).
The incident pump has a spectrum that is centered at $f_0 = \omega_0/2\pi = 375\ \mathrm{THz}$, and contains two frequency components $\omega_1 = 1.05\omega_0$ and $\omega_2 = 0.95\omega_0$ that are phase-matched for $B_c$ and $N_0$.
The transmitted spectrum exhibits a new spectral peak centered at $\omega_3 = \omega_1 - \omega_2 = 0.1\omega_0$, corresponding to $f_3 = \omega_3/2\pi = 37.5\ \mathrm{THz}$. 
This terahertz output is a single-cycle pulse with a peak electric field of 440 GV/m, containing 5\% of the incident energy.

The high field strengths and efficiencies are a direct consequence of phase-matched wave mixing in magnetized plasma.
For the frequency conversion process, the phase mismatch is defined as $\Delta k = k_{1} - k_{2} - k_{3}$, where $k_i$ denotes the wavenumber at frequency $\omega_i$ $(i = 1, 2, 3)$. % 
When the external magnetic field is oriented orthogonal to both the pump polarization and the propagation direction, only X modes are excited in the plasma, for which the dispersion relation is:
\begin{equation}
    \label{eq:dispersion relation of X modes}
    n_\mathrm{X, \omega}^2 = \frac{c^2k^2}{\omega^2} = 1 - \frac{\omega_p^2}{\omega^2}\frac{\omega^2-\omega_p^2}{\omega^2 - \omega_H^2},
\end{equation}
where $n_\mathrm{X, \omega}$ is the refractive index of X modes at frequency $\omega$, $\omega_p = \sqrt{n_ee^2/(\epsilon_0m_e)}$ is plasma frequency, $n_e$ is plasma density, $\epsilon_0$ is the vacuum permittivity, $\omega_H = (\omega_p^2 + \omega_c^2)^{1/2}$ is the upper hybrid frequency, $\omega_c = eB_0/m_e$ is the electron cyclotron frequency, and $B_0$ is the external magnetic field.
The X-mode dispersion relation supports two bands of propagating modes in the plasma, which can be exploited for phase-matched wave mixing. 
Phase-matching is achieved by tuning the plasma density and magnetic field strength such that $\omega_3$ lies on the lower X-mode branch, $\omega_1$ and $\omega_2$ lie on the upper branch, and the condition $k_3 = k_1 - k_2$ is satisfied. 
A representative phase-matched dispersion diagram is shown in Fig.~\ref{fig:THz_schematic}(b). 
The phase-matching condition is:
\begin{equation}
    \label{eq:phase matching condition}
    n_{\mathrm{X},\omega_1}\omega_1 - n_{\mathrm{X},\omega_2}\omega_2 - n_{\mathrm{X},\omega_3}\omega_3 = 0.
\end{equation}
Defining $\omega_3 = \omega_1 - \omega_2 = 2\alpha\omega_0$ and $\omega_0 = (\omega_1 + \omega_2)/2$, the phase-matching condition can be rewritten as
\begin{equation}
    \label{eq:phase matching condition normalized units}
    f(1 + \alpha) - f(1 - \alpha) = f(2\alpha),
\end{equation}
where
\begin{equation}
    \label{eq:fx}
    f(x) = \sqrt{x^2 - N_0\frac{x^2 - N_0}{x^2 - N_0 - B_c^2}}
\end{equation}
with $N_0 = n_e/n_c$ the normalized plasma density, $n_c = \epsilon_0m_e\omega_0^2/e^2$ the critical density at $\omega_0$, and $B_c = \omega_c/\omega_0 = eB_0/(m_e\omega_0)$ the normalized magnetic field strength.
The terahertz frequency is thus tunable through $\omega_0$ and $\alpha$: for instance, $\alpha = 0.1$ and $f_0 = \omega_0/2\pi = 375\ \mathrm{THz}$ yields $f_3 =0.2f_0 =  75\ \mathrm{THz}$, while $\alpha = 0.05$ and $f_0 = 30\ \mathrm{THz}$ yields $f_3 = 0.1f_0 = 3\ \mathrm{THz}$.
For any combination of $\omega_0$ and $\alpha$, the required plasma density and magnetic field strength for phase matching are determined from Eqs.~\eqref{eq:phase matching condition}--\eqref{eq:fx}.

The THz generation process and phase-matching conditions were validated using the PIC code EPOCH~\cite{arber2015contemporary}. All simulations employed a resolution of 80 cells/$\lambda_0$ with 20 particles per cell, which was verified to be sufficient for numerical convergence by comparison with higher-resolution runs.

\begin{figure}[t]
    \centering
    \includegraphics[width=1.0\linewidth]{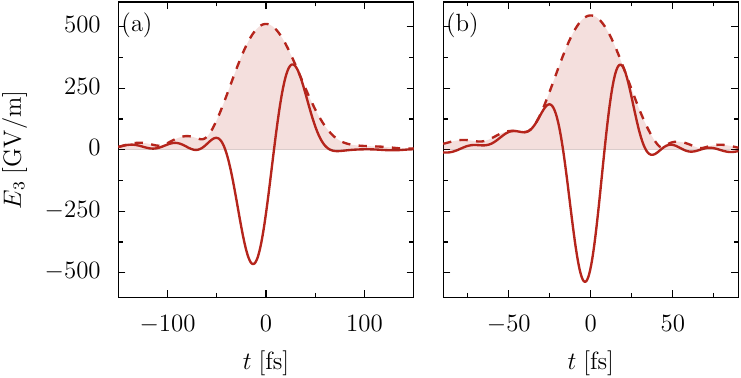}
    \caption{PIC simulation results of generated terahertz pulses with field strengths greater than $500\ \mathrm{GV/m}$. 
    (a) The terahertz pulse generated by sending a single broadband pulse pulse with $f_0 = 375\ \mathrm{THz}$, $\tau_0 = 48\ \mathrm{fs}$, $a_0 = 2.3$, through a magnetized plasma with $B_c = 0.39$, $N_0 = 3.16\times 10^{-4}$, and $L = 2800\lambda_0$. 
    (b) The terahertz pulse generated by sending a single broadband pulse pulse with $f_0 = 375\ \mathrm{THz}$, $\tau_0 = 24\ \mathrm{fs}$, $a_0 = 1.4$, through a magnetized plasma with $B_c = 0.36$, $N_0 = 0.02$, and $L = 500\lambda_0$. }
    \label{fig:THz_examples}
\end{figure}

\begin{figure}[t]
    \centering
    \includegraphics[width=1\linewidth]{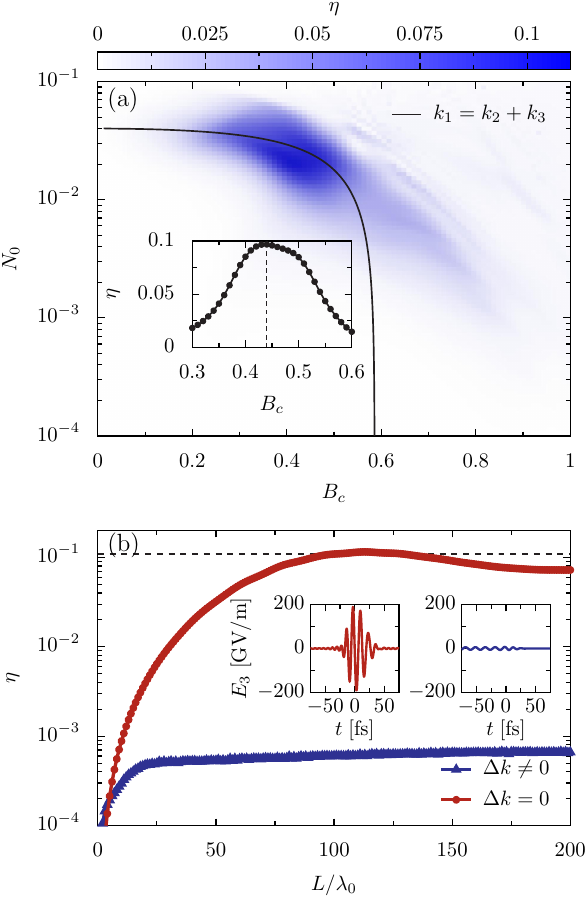}
    \caption{(a) Conversion efficiency scan of THz generation with $\alpha = 0.1$ at varied plasma density $N_0$ and magnetic field strength $B_c$. The plasma length is $L = 100\lambda_0$. The solid line marks the phase-matching condition and the inset shows the efficiency versus magnetic field at $N_0 = 0.025$ with the dashed line marks the magnetic field for phase-matching. (b) Conversion efficiency at varied plasma lengths for phase-matched ($B_c = 0.44$ and $N_0 = 0.025$) and phase-mismatched ($B_c = 0.1$ and $N_0 = 0.025$) THz generation. The inset shows the terahertz field for both cases when $L = 100\lambda_0$.}
    \label{fig:1d_PIC_scan}
\end{figure}

Frequency tunability is achieved by varying the pump source and adjusting the plasma density and magnetic field strength to select which spectral components of the pump are phase-matched.
Figure~\ref{fig:THz_examples} shows two representative examples, both yielding output field strengths near 500 GV/m at different terahertz frequencies. 
In Fig.~\ref{fig:THz_examples}(a), a single-cycle Gaussian terahertz pulse is generated at a central frequency $f_3 = 0.02\omega_0/2\pi = 7.5\ \mathrm{THz}$, driven by phase-matched pump components at $\omega_1 = 1.01\omega_0$ and $\omega_2 = 0.99\omega_0$, with a conversion efficiency of 0.5\%. 
Figure~\ref{fig:THz_examples}(b) shows a second single-cycle pulse at $f_3 = 0.05\omega_0/2\pi = 18.75\ \mathrm{THz}$, generated by components at $\omega_1 = 1.025\omega_0$ and $\omega_2 = 0.975\omega_0$, achieving a peak field of 550 GV/m and a conversion efficiency of 1.8\%. 
The lower efficiency in the first case is consistent with the larger photon energy difference associated with the lower terahertz frequency. 
In both cases, the generated field strengths correspond to $a_0 > 2$, sufficient to drive electrons to relativistic quiver velocities. 

Figure~\ref{fig:1d_PIC_scan}(a) shows the conversion efficiency as a function of plasma density and magnetic field strength for a two-color pump pulse with $\alpha = 0.1$, $f_0 = 375\ \mathrm{THz}$, duration $\tau_{1,2} = 35\ \mathrm{fs}$ full-width-at-half-maximum (FWHM), and normalized vector potential $a_1 = a_2 = eE_{1,2}/(m_e\omega_{1,2}c) = 0.1$, where $E_{1,2}$ is the peak electric field amplitude of each color.
The generated terahertz radiation has a frequency $f_3 = 0.2f_0 = 75\ \mathrm{THz}$. 
The conversion efficiency was computed from the energy spectrum of the Fourier-transformed transmitted pulse.
The efficiency peaks near the phase-matching condition of Eq.~\eqref{eq:phase matching condition}, with a maximum of approximately 10.6\% at $B_c \approx 0.44$ and $N_0\approx 0.025$ after a propagation distance of $100\lambda_0$.
Away from the maximum, efficiency decreases for two reasons: decreasing magnetic field strength narrows the lower X-mode propagation band, eventually prohibiting terahertz wave propagation, while decreasing plasma density weakens the coupling between the electromagnetic and density waves.

The theoretical maximum conversion efficiency, obtained by assuming complete depletion of $\omega_1$ photons into $\omega_2$ and $\omega_3$ photons, is $\eta_\mathrm{max} = N_1\omega_3/(N_1\omega_1+N_2\omega_2)$, where $N_1$ and $N_2$ are the input photon number density of $\omega_1$ and $\omega_2$, respectively. 
For the parameters considered here, this yields $\eta_\mathrm{max}=10.9\%$, in agreement with the simulated value of 10.6\%, suggesting that the process operates near the photon efficiency limit. 
The inset of Fig.~\ref{fig:1d_PIC_scan}(a) shows the efficiency as a function of magnetic field strength at fixed plasma density, revealing a peak centerd at the phase-matched value with a width of $\Delta B_c\approx 0.15$.
This tolerance to magnetic field variations indicates that the phase-matching condition is robust against perturbations and that the process supports broadband pulses with correspondingly short pulse durations.

Figure~\ref{fig:1d_PIC_scan}(b) compares the conversion efficiency of phase-matched and phase-mismatched THz generation as a function of interaction length, with the same plasma densities but different magnetic field strength in the two cases.
In the phase-matched case, the conversion efficiency grows monotonically with interaction length, saturating around 10\% at a propagation distance of 100$\lambda_0$. 
The saturation arises because the growing THz wave, whose frequency is near the plasma frequency, begins to strongly interact with the plasma and modify the conversion process.
In the phase-mismatched case, by contrast, the efficiency is more than two orders of magnitude lower and remains nearly constant with increasing interaction length. % than  
The insets of Fig.~\ref{fig:1d_PIC_scan}(b) show the output terahertz field in both cases at $L = 100\lambda_0$.
The phase-matched case produces a multi-cycle Gaussian terahertz pulse at $f_3 = 0.2f_0 = 75\ \mathrm{THz}$ with a peak field strength around $200\ \mathrm{GV/m}$, while the phase-mismatched case yields a substantially weaker signal.
This comparison underscores the significance of phase matching in achieving high conversion efficiency and ultrahigh terahertz field strengths.

\begin{figure}[t]
    \centering
    \includegraphics[width=1.0\linewidth]{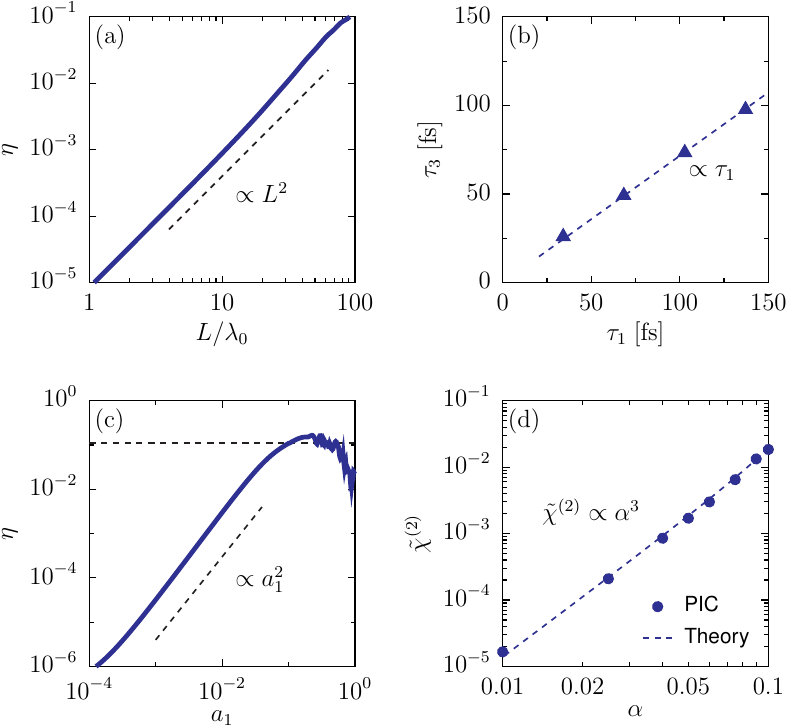}
    \caption{(a) The conversion efficiency to THz radiation for varied propagation distance. (b) The FWHM of the generated THz pulse as a function of the FWHM of the pump pulse. (c) The conversion efficiency of THz radiation at varied pump strength $a_1$ and fixed $a_2 = 0.1$. (d) The nonlinear strength of the generation process at varied frequencies $\omega_3$, and fixed $B_0$ and $\omega_3^2/\omega_0^2$.}
    \label{fig:THz_scaling}
\end{figure}

\begin{figure}[t]
    \centering
    \includegraphics[width=1\linewidth]{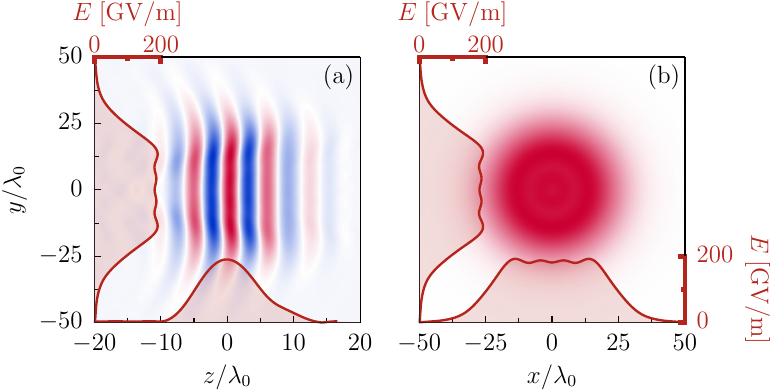}
    \caption{Three-dimensional PIC simulation of THz generation by sending a two-color pump pulse with $a_{1,2} = 0.1$, $\tau_{1,2} = 35\ \mathrm{fs}$, $\alpha = 0.1$, and $f_0 = 375\ \mathrm{THz}$ through a magnetized plasma with $B_c = 0.4$, $N_0 = 0.025$, and $L = 100\lambda_0$. 
    (a) The longitudinal oscillating electric field and its envelopes of the output terahertz pulse. 
    (b) The transverse envelopes of the output terahertz pulse.}
    \label{fig:3d_PIC}
\end{figure}

The physics underlying the terahertz generation process can be understood by analyzing the electron motion in the magnetized plasma with the presence of a two-color pump wave. 
Solving the Lorentz equation to second order yields the effective nonlinearity, which in normalized units is~\cite{Supplemental_Material}
\begin{equation}
    \label{eq:nonlinearity}
    \left|\tilde{\chi}^{(2)}\right| \approx\frac{\alpha N_0B_c(1 + 3B_c^2)}{(4\alpha^2 - B_c^2)(1 - N_0 - B_c^2)^2}.
\end{equation}
Since the normalized plasma density is approximately constant across different output frequencies $\omega_3$~\cite{Supplemental_Material}, the nonlinearity scales as 
\begin{equation}
    \label{eq:nonlinearity scaling}
    N_0 \propto \text{constant}\times \alpha^2\Rightarrow\left|\tilde{\chi}^{(2)}\right|\propto (\omega_3/\omega_0)^3.
\end{equation}
Given the input pump amplitudes and propagation distance, the output terahertz field strength can be determined from Eq.~\eqref{eq:nonlinearity}~\cite{Supplemental_Material}, yielding a conversion efficiency
\begin{equation}
    \label{eq:efficiency}
    \eta = \frac{E_3^2}{E_{10}^2 + E_{20}^2}\propto\frac{(\chi^{(2)}a_1a_2L\omega_3)^2}{\omega_1^2a_1^2 + \omega_2^2a_2^2}\propto L^2\omega_3^6, 
\end{equation}
where $E_{10}$ and $E_{20}$ are the peak electric field amplitudes of the two pump colors at frequencies $\omega_1$ and $\omega_2$ respectively, and $\chi^{(2)} = e\tilde{\chi}^{(2)}/(m_e\omega_3c)$ is the nonlinear coupling strength.
Full derivations of the nonlinearity and terahertz field strength are provided in the supplemental material~\cite{Supplemental_Material}. 

Equations~\eqref{eq:nonlinearity} to \eqref{eq:efficiency} can be used to interpret the features of the terahertz generation process. 
Figure~\ref{fig:THz_scaling}(a) confirms that the conversion efficiency scales quadratically with propagation distance in the phase-matched case $\eta\propto L^2$, consistent with Eq.~\eqref{eq:efficiency}.
Figure~\ref{fig:THz_scaling}(b) shows that the terahertz pulse duration is proportional to, and shorter than, the pump duration, and can therefore be controlled by adjusting the pump pulse duration. 
Figure~\ref{fig:THz_scaling}(c) shows that when the efficiency is low and one pump is significantly stronger than the other, the conversion efficiency grows quadratically with pump field strength. 
The efficiency saturates as it approaches $\eta_\mathrm{max}$, and can exceed this estimate due to secondary wave mixing processes in which $\omega_2$ photons are converted to $\omega_3$ photons via $\omega_2\to \omega_3 + \omega_4$ (where $\omega_4 = \omega_2 - \omega_3$).
At sufficiently high pump strengths, other plasma nonlinearities begin to affect terahertz growth, causing the efficiency to decrease. 
Figure~\ref{fig:THz_scaling}(d) shows the normalized nonlinearity extracted from the output terahertz field amplitude at different terahertz frequencies, which scales as $\tilde{\chi}^{(2)}\propto (\omega_3/\omega_0)^{3}$, in close agreement with the theoretical prediction of Eq.~\eqref{eq:nonlinearity scaling}. 
This frequency scaling implies that generating lower-frequency terahertz radiation requires higher pump intensities and longer interaction lengths to achieve a high field strength.

A three-dimensional PIC simulation was performed to assess the impact of multidimensional effects on the generation process. 
The two-color pump pulse had a super-Gaussian spatial profile of order 10 and $\sigma = 24\ \mathrm{\upmu m}$, while the temporal profile and input field strengths were identical to those used in Fig.~\ref{fig:1d_PIC_scan}(a). 
As shown in Fig.~\ref{fig:3d_PIC}(a) and (b), the generated terahertz pulse exhibits a Gaussian temporal profile and a nearly flat spatial profile with a transverse size closely matching that of the pump.
The peak electric field of 180 $\mathrm{GV/m}$ is in good agreement with the one-dimensional result (inset of Fig.~\ref{fig:1d_PIC_scan}(b)), confirming that multidimensional effects have negligible influence on the output field strength.  

Compared to THz generation from laser wakefield in strongly magnetized plasmas~\cite{tailliez2022terahertz,tailliez2023terahertz}, the phase-matched DFG process offers higher conversion efficiency and greater flexibility in controlling terahertz frequency and pulse duration. 
Additionally, in contrast to the optimal condition for terahertz generation from laser wakefield, which requires the ratio between the electron cyclotron frequency and the relativistic plasma frequency to be slightly greater than unity~\cite{tailliez2022terahertz}, the phase-matched DFG process requires the electron cyclotron frequency to be much greater than the relativistic plasma frequency.
These distinctions not only position phase-matched DFG in a complementary parameter space but also relax pump intensity requirements for producing ultrahigh THz field strengths. 
Our results thus expand the landscape of intense THz generation and suggest new pathways toward high-field THz sources.

In conclusion, we have demonstrated both theoretically and numerically that phase-matched difference frequency generation in strongly magnetized plasma can produce terahertz pulses with tunable frequency and pulse duration, and peak field strengths exceeding hundreds of $\mathrm{GV/m}$ driven by either a two-color pump or a spectrally broad short pulse.
Analytic expressions for the phase-matching conditions and the nonlinear interaction strength are derived and shown to be in excellent agreement with simulation results.
These findings establish a new mechanism for intense terahertz generation that exploits the two branches of X modes to achieve high efficiency and ultrahigh field strengths, extending high-field THz science into the regime of extreme plasma physics.

\medskip

\noindent {\bf Acknowledgments. }This work was partially supported by NSF Grant PHY-2308641 and NNSA Grant DE-NA0004130. The PIC code EPOCH~\cite{arber2015contemporary} is funded by the UK EPSRC grants EP/G054950/1, EP/G056803/1, EP/G055165/1 and EP/ M022463/1. The computing for this project was performed on the (Stanford) Sherlock cluster. We would like to thank Stanford University and the Stanford Research Computing Center for providing computational resources and support that contributed to these research results.

\noindent {\bf Disclosures.} The authors declare no conflicts of interest.

\noindent {\bf Data availability. }Data underlying the results presented in this paper are not publicly available at this time but may be obtained from the authors upon reasonable request.

\bibliography{reference}% Produces the bibliography via BibTeX.

\onecolumngrid
\newpage

\section{Supplemental Material}

\subsection{Derivation of the nonlinear interaction strength}
To understand the terahertz generation process, we start by considering the electron motion in the presence of the external magnetic field and the two-color pump wave using the fluid equation and the Lorentz equations, assuming the plasma is cold and collisionless 
\begin{subequations}
\label{eq:fluid equation and lorentz equation}
\begin{eqnarray}
       \frac{\partial \mathbf{v}}{\partial t} + (\mathbf{v}\cdot\mathbf{\nabla})\mathbf{v} &=& -\frac{e}{m_e}\left(\mathbf{E} + \frac{\mathbf{v}}{c}\times \mathbf{B}\right)\\
       \frac{\partial n_e}{\partial t} + \mathbf{\nabla}\cdot(n_e\mathbf{v}) &=& 0.
\end{eqnarray}
\end{subequations}
We expand the electron velocities, density perturbations and magnetic fields using the perturbation series $\mathbf{v} = \sum_{i,j}\mathbf{v}_{\omega_j}^{(i)}$, $n_{e,\omega_j} = n_{e0} + \sum_{i,j}n_{e,\omega_j}^{(i)}$, and $\mathbf{B} = \mathbf{B}_0 +\sum_{i,j}\mathbf{B}_{\omega_j}^{(i)}$, where the summation over $j = 1,2$ accounts for the contribution from the two pumps at different frequencies and the summation over $i$ denotes different orders. 
We first solve for the first-order response $n_{e,\omega_j}^{(1)}$ and $\mathbf{v}_{\omega_j}^{(1)}$, which is the solution to the extraordinary mode. 
We let the pump pulse propagate in the $z$ direction, with the electric field in $y$ and magnetic field in $x$. 
The external magnetic field $\mathbf{B}_0 = B_0\hat{x}$. 
In the one-dimensional case, the first-order quantities are given by
\begin{subequations}
\label{eq:first_order}
\begin{eqnarray}
        v_{z,\omega_j}^{(1)} &=& -\frac{a_jc\omega_c\omega_j}{\omega_j^2 - \omega_H^2}\\
        v_{y,\omega_j}^{(1)} &=& -i\frac{a_jc(\omega_j^2-\omega_p^2)}{\omega_j^2 - \omega_H^2}\\
        n_{\omega_j}^{(1)} &=& -\frac{n_{e}a_jk_jc\omega_c}{\omega_j^2 - \omega_H^2},
\end{eqnarray}
\end{subequations}
where $\omega_c = eB_0/m_e$ is the cyclotron frequency, $\omega_H^2 = \omega_p^2 + \omega_c^2$ is the upper hybrid frequency, $\omega_p^2 = n_{e0}e^2/(\epsilon_0m_e)$ is the squared plasma frequency, $a_j = eE_j/(m_e\omega_jc)$ is the normalized vector potential corresponding to the pump wave at frequency $\omega_j$, $E_j$ is its electric field amplitude, $c$ is the speed of light, and $k_j$ is the wavenumber of an extraordinary wave at frequency $\omega_j$. 
These quantities given in Eq.~\eqref{eq:first_order} are validated using PIC simulations. 
Next, we consider the second-order Lorentz equations for electron motion at frequency $\omega_3$
\begin{subequations}
\label{eq:second order}
\begin{eqnarray}
        \frac{\partial v_{y,\omega_3}^{(2)}}{\partial t} - \frac{e}{m_e}v_{z,\omega_3}^{(2)}B_0 &=& -\left(v_{z,\omega_1}^{(1)}\frac{\partial v_{y,\omega_2}^{(1),*}}{\partial z} + v_{z,\omega_2}^{(1),*}\frac{\partial v_{y,\omega_1}^{(1)}}{\partial z}\right) + \frac{e}{m_ec}\left(v_{z,\omega_1}^{(1)}B_{x,\omega_2}^{(1),*} + v_{z,\omega_2}^{(1),*}B_{x,\omega_1}^{(1)}\right)\\
        \frac{\partial v_{z,\omega_3}^{(2)}}{\partial t} + \frac{e}{m_e}v_{y,\omega_3}^{(2)}B_0 &=& -\left(v_{z,\omega_1}^{(1)}\frac{\partial v_{z,\omega_2}^{(1),*}}{\partial z} + v_{z,\omega_2}^{(1),*}\frac{\partial v_{z,\omega_1}^{(1)}}{\partial z}\right) - \frac{e}{m_ec}\left(v_{y,\omega_1}^{(1)}B_{x,\omega_2}^{(1),*} + v_{y,\omega_2}^{(1),*}B_{x,\omega_1}^{(1)}\right).
\end{eqnarray}
\end{subequations}
We simplify Eq.~\eqref{eq:second order} by substituting $\partial/\partial t\to -i\omega_j$ and $\partial/\partial z\to ik_j$, then Eq.\eqref{eq:second order} becomes
\begin{subequations}
\label{eq:second order nonlinear equations}
\begin{eqnarray}
        F_y \equiv-i\omega_3v_{y,\omega_3}^{(2)} - \omega_cv_{z,\omega_3}^{(2)} &=& -i(k_1v_{y,\omega_1}^{(1)}v_{z,\omega_2}^{(1),*} - k_2v_{y,\omega_2}^{(1),*}v_{z,\omega_1}^{(1)}) + a_2^*\omega_2v_{z,\omega_1}^{(1)} + a_1\omega_1v_{z,\omega_2}^{(1),*}\\
        F_z\equiv-i\omega_3v_{z,\omega_3}^{(2)} + \omega_cv_{y,\omega_3}^{(2)} &=& -i(k_1v_{z,\omega_1}^{(1)}v_{z,\omega_2}^{(1),*} - k_2v_{z,\omega_2}^{(1),*}v_{z,\omega_1}^{(1)}) - a_2^*\omega_2v_{y,\omega_1}^{(1)} - a_1\omega_1v_{y,\omega_2}^{(1),*}.
\end{eqnarray}
\end{subequations}
Solving for $v_{y,\omega_3}^{(2)}$, we find
\begin{equation}
    \label{eq:second order velocity}
    v_{y,\omega_3}^{(2)} = \frac{-\omega_cF_z + i\omega_3F_y}{\omega_3^2 - \omega_c^2}.
\end{equation}
This non-zero velocity component at frequency $\omega_3$ is a result of the beating of the two pump waves, not a result of the generated electric field at frequency $\omega_3$.
We can determine the nonlinear current that contributes to the terahertz generation at frequency $\omega_3$ using Eq.~\eqref{eq:first_order} and Eq.~\eqref{eq:second order velocity}
\begin{equation}
    \label{eq:nonlinear current}
    J_{y,\omega_3} = -\frac{1}{2}e\left(n_{e0}v_{y,\omega_3}^{(2)} + n_{\omega_1}^{(1)}v_{y,\omega_2}^{(1),*} + n_{\omega_2}^{(1),*}v_{y,\omega_1}^{(1)}\right).
\end{equation}
Plugging Eq.~\eqref{eq:first_order} and Eq.~\eqref{eq:second order velocity} into Eq.~\eqref{eq:nonlinear current}, we can find the nonlinear current contributing to the THz generation, which is general and regardless of the phase-mismatch. 
We divide Eq.~\eqref{eq:nonlinear current} into two terms $J_1 = -e((n_{e0}v_{y,\omega_3}^{(2)})/2$ and $J_2 = -e(n_{\omega_1}^{(1)}v_{y,\omega_2}^{(1),*} + n_{\omega_2}^{(1),*}v_{y,\omega_1}^{(1)})/2$, plug in the first and second order quantities, and simplify using the phase-matching conditions ($\omega_1 - \omega_2 = \omega_3$ and $n_1\omega_1- n_2\omega_2 = n_3\omega_3$), then the explicit expressions for the current are: 
\begin{subequations}
\label{eq:explicit current expressions}
\begin{eqnarray}
        J_1 & \approx & -i\frac{c\epsilon_0m_e\omega_p^2\omega_c\omega_3}{2e(\omega_3^2 - \omega_c^2)}\frac{\left[(\omega_1^2 - \omega_p^2)(\omega_2^2 - \omega_p^2) + 4\omega_1\omega_2\omega_c^2\right]}{(\omega_1^2-\omega_H^2)(\omega_2^2 - \omega_H^2)}a_1a_2 + O(\omega_3^5)\\
        J_2 & = & i\frac{c\epsilon_0m_e\omega_p^2\omega_c\omega_3\left[\omega_1\omega_2(n_3 - n_1 - n_2) - n_3\omega_p^2\right]}{2e(\omega_1^2-\omega_H^2)(\omega_2^2 - \omega_H^2)}a_1a_2 \approx -i\frac{c\epsilon_0m_e\omega_p^2\omega_c\omega_1\omega_2\omega_3}{2e(\omega_1^2-\omega_H^2)(\omega_2^2 - \omega_H^2)}a_1a_2 + O(\omega_3^5).
\end{eqnarray}
\end{subequations}
The total nonlinear current is the sum of Eqs.~\eqref{eq:explicit current expressions}, 
\begin{equation}
    \label{eq:total nonlinear current explicit expression}
    J_{y,\omega_3} \approx -i\frac{c\epsilon_0m_e\omega_p^2\omega_c\omega_3}{2e(\omega_3^2 - \omega_c^2)}\frac{\left[(\omega_1^2 - \omega_p^2)(\omega_2^2 - \omega_p^2) + 3\omega_1\omega_2\omega_c^2\right]}{(\omega_1^2-\omega_H^2)(\omega_2^2 - \omega_H^2)}a_1a_2 
\end{equation}
Then the normalized nonlinear strength of this process can be determined as
\begin{equation}
    \label{eq:nonlinearity}
    \left|\tilde{\chi}^{(2)}\right| = \left|\chi^{(2)}\right|\left[\frac{m_e\omega_3c}{e}\right] = \left|\frac{J_{y,\omega_3}}{\omega_3\epsilon_0E_{10}E_{20}} \right|\left[\frac{m_e\omega_3c}{e}\right]\approx \frac{\omega_p^2\omega_c\omega_3}{2(\omega_3^2 - \omega_c^2)}\frac{\omega_1\omega_2 + 3\omega_c^2}{(\omega_1^2-\omega_H^2)(\omega_2^2 - \omega_H^2)},
\end{equation}
and in terms of normalized units
\begin{equation}
    \label{eq:normalized nonlinearity}
    \left|\tilde{\chi}^{(2)}\right| \approx\frac{\alpha N_0B_c(1 + 3B_c^2)}{(4\alpha^2 - B_c^2)(1 - N_0 - B_c^2)^2}.
\end{equation}
Using the nonlinearity given by Eq.~\eqref{eq:nonlinearity}, we can calculate the the field strength of the output terahertz wave
\begin{equation}
\label{eq:terahertz amplitude}
|E_3| =  \left|\frac{\omega_3\chi^{(2)}}{2n_{\mathrm{X},\omega_3}c}E_{10}E_{20}L\right| \approx   \left|\frac{m_e\omega_p^2\omega_c\omega_1\omega_2\omega_3}{4e(\omega_3^2 - \omega_c^2)}\frac{\omega_1\omega_2 + 3\omega_c^2}{(\omega_1^2-\omega_H^2)(\omega_2^2 - \omega_H^2)}a_1a_2L\right|.
\end{equation}

\begin{figure}[t]
    \centering
    \includegraphics[width=1.0\linewidth]{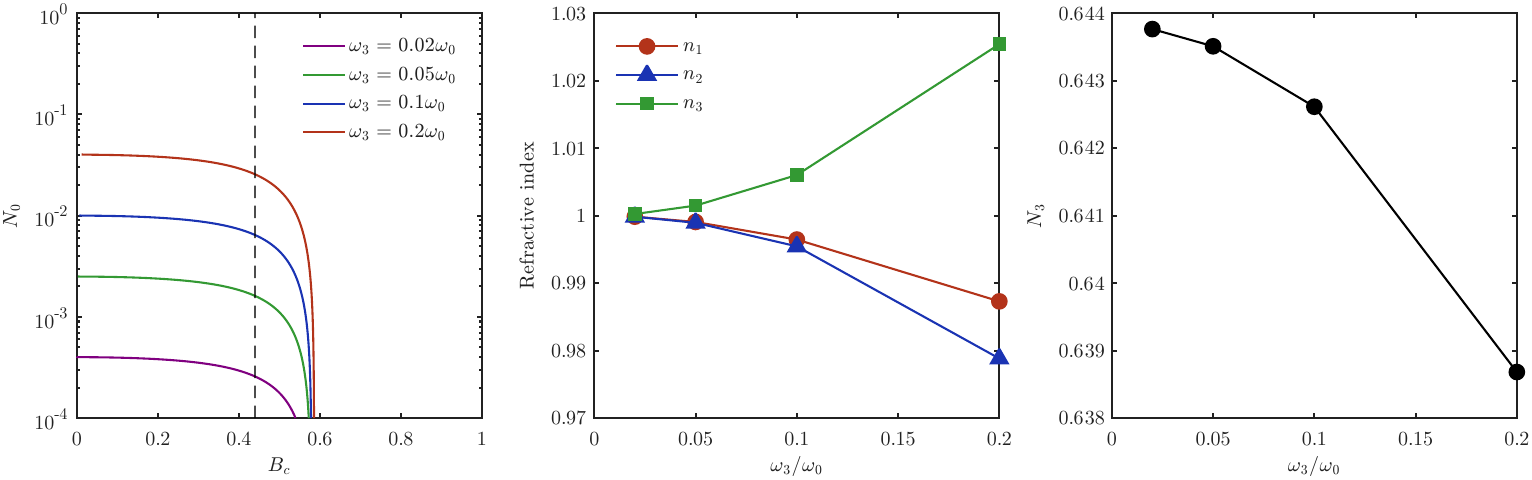}
    \caption{(a) Phase-matching conditions as a function of $N_0$ and $B_c$ at different THz frequency $\omega_3$. 
    (b) The refractive indices of three waves at frequency $\omega_1$, $\omega_2$, and $\omega_3$ at different $\omega_3$ frequency.
    (c) The normalized plasma density at frequency $\omega_3$ at fixed $B_c = 0.44$.}
    \label{fig:Phase-matching conditions}
\end{figure}

To derive Eq.~\eqref{eq:explicit current expressions} to Eq.~\eqref{eq:terahertz amplitude}, we used several approximations, including $n_{\mathrm{X,\omega_1}}\approx n_{\mathrm{X,\omega_2}}\approx n_{\mathrm{X,\omega_3}} = 1$, $\alpha^2\ll 1$, $\omega_1^2 - \omega_p^2\approx \omega_1^2$, and $\omega_2^2-\omega_p^2 \approx \omega_2^2$.
These relations are justified by Fig.~\ref{fig:Phase-matching conditions}. 
As shown in Fig.~\ref{fig:Phase-matching conditions}(a), for different output THz frequency, the phase-matching condition requires tenuous plasma $N\ll 1$. 
Therefore, we simplify Eq.~\eqref{eq:normalized nonlinearity} using $\omega_1^2 - \omega_p^2\approx \omega_1^2$ and $\omega_2^2-\omega_p^2 \approx \omega_2^2$. 
Figure~\ref{fig:Phase-matching conditions}(b) shows the refractive indices of the three waves at a fixed $B_c$ but different plasma densities due to phase-matching at different output THz frequencies.
As the output frequency varies from $0.02\omega_0$ to $0.2\omega_0$, the indices of the three waves are all around 1 and have a maximum fluctuation about $3\%$. 
Therefore, we approximated $n_{\mathrm{X,\omega_1}}\approx n_{\mathrm{X,\omega_2}}\approx n_{\mathrm{X,\omega_3}} = 1$ to simplify Eq.~\eqref{eq:normalized nonlinearity}. 
Figure~\ref{fig:Phase-matching conditions}(c) shows that the normalized plasma density at the output frequency $N_3 = \omega_p^2 / \omega_3^2$ is approximately a constant on the order of unity. 
Therefore, we can find how the nonlinearity and the output field strength scale with the output frequency
\begin{subequations}
\label{eq:normalized nonlinearity scaling}
\begin{eqnarray}
        \left|\tilde{\chi}^{(2)}\right| &\approx&\frac{4\alpha^3 N_3B_c(1 + 3B_c^2)}{(4\alpha^2 - B_c^2)(1 - N_0 - B_c^2)^2}\propto \alpha^3\propto (\omega_3/\omega_0)^3\\
        |E_3| &\approx&   \left|\frac{m_e\omega_p^2\omega_c\omega_1\omega_2\omega_3}{4e(\omega_3^2 - \omega_c^2)}\frac{\omega_1\omega_2 + 3\omega_c^2}{(\omega_1^2-\omega_H^2)(\omega_2^2 - \omega_H^2)}a_1a_2L\right|\propto (\omega_3/\omega_0)^3a_1a_2L.
\end{eqnarray}
\end{subequations}
The scaling given by Eq.~\eqref{eq:normalized nonlinearity scaling} is used to find the theoretical nonlinear strength shown in Fig.~4(d) in the manuscript, which agrees reasonably well with PIC simulation results.

\subsection{Detailed Computational Parameters}
\begin{table}[h]
\caption{Physical and Computational Parameters}
\label{tbl:params}
\begin{ruledtabular}
\begin{tabular}{l c c c c c c c c c c c c}
\noalign{\smallskip}
{\bf Figure} &
Code\footnote{For further information on the PIC codes used, see~\cite{arber2015contemporary} (EPOCH).} &
 Dim.\footnote{All simulations calculate all three components of particle velocity.} &
  Cells/$\lambda_0$\footnote{The number of grid cells per wavelength $\lambda_0$, where $\lambda_0 = 2\pi c/\omega_0$, and $\omega_0 = (\omega_1 + \omega_2) /2$.}&
Part./Cell\footnote{Number of particles in a cell in regions where plasma density is non-zero at the beginning of the simulation.}&
$N_0$\footnote{$N_0 = n_e/n_c$ where $n_e$ is the electron number density and $n_c$ is the critical density for light at wavelength $\lambda_0$ ($n_c = \epsilon_0 m_e \omega_0^2 / e^2$).} &
$B_c$\footnote{$B_c = eB_0/m_e\omega_0$, where $B_0$ is applied magnetic field and is transverse to the laser propagation direction.} & 
 
$a_{0}$ \footnote{$a_{0} = eE_{0}/m_e\omega_0c$, where $E_{0}$ is the maximum electric field of the laser pulse with a central frequency $\omega_0$. The electric field is polarized perpendicular to the magnetic field.} & $a_{1,2}$ \footnote{$a_{1,2} = eE_{1,2}/m_e\omega_{1,2}c$, where $E_{1,2}$ is the maximum electric field of the laser pulse with a central frequency $\omega_{1,2}$.}& $\tau_0$\footnote{All pulses are Gaussian and $\tau$ is the full-width-half-maximum (FWHM) duration, such that:
\begin{equation}
I(t) = I_0 e^{-\left(\frac{t-t_0}{\tau / 2\sqrt{\ln 2}}\right)^2}
\end{equation}} &
$\tau_{1,2}$&
$L/\lambda_0$\footnote{Plasma length normalized by the wavelength $\lambda_0 = 2\pi c/\omega_0$.} \\ 
%%%%%
\noalign{\smallskip}
\hline
\noalign{\medskip}
1(a) & EPOCH 	& 1 & 80 & 20 &  0.01 & 0.4 & 0.6 & - & 12 fs& - & 100 \\
2(a) & EPOCH & 1 & 80 & 20 &  $3.16\times 10^{-4}$ & 0.39 & 2.3 & - & 48 fs & - & 2800 \\ 
2(b) & EPOCH & 1 & 80 & 20 & 0.02 & 0.36 & 1.4 & - & 24 fs & -  & 500 \\
3(a) & EPOCH & 1 & 80 & 20 & [$10^{-4}$-$10^{-1}$] & $[0.1$-$1$] & -  & 0.1 & - & 35 fs & 100   \\
3(b) & EPOCH & 1 & 80 & 20 & 0.025 & 0.44, 0.1 & - & 0.1 & - & 35 fs & [$1$-$100$]  \\
4(a) & EPOCH & 1 & 80 & 20 & 0.025 & 0.44 & - & 0.1 & - & 35 fs	&  [$1$-$100$] \\
4(b) & EPOCH & 1 & 80 & 20 & 0.025 & 0.44 & - & 0.1 & - & 35-140 fs	& 100 \\
4(c) & EPOCH & 1 & 80 & 20 & 0.025 & 0.44 & - & [$10^{-4}$-$10^0$] & - & 35 fs	& 100 \\
4(d) & EPOCH & 1 & 80 & 20 & [0.00025-0.025] & 0.44 & - & 0.1 & - & 140 fs	& 100  \\
5 & EPOCH & 3 & 12 & 1 & 0.025 & 0.4 & - & 0.1 & - & 35 fs & 100  \\
\end{tabular}
\end{ruledtabular}
\end{table}

\end{document}